\documentclass[pre]{revtex4}
\usepackage{amsmath}
\usepackage{graphicx}
\begin{document}
\title{Numerical Simulation of Three-Dimensional Dendrites using Coupled Map Lattices}
\author{Hidetsugu Sakaguchi, Mai Gondou, and Haruo Honjo}
\affiliation{Department of Applied Science for Electronics and Materials, \\
Interdisciplinary Graduate School of Engineering Sciences, Kyushu
University, \\
Kasuga, Fukuoka 816-8580, Japan}
\begin{abstract}
Three dimensional dendrites are studied with a coupled map lattice model. 
We study the fractal dimensions, the $f(\alpha)$ spectrum, the size distribution of sidebranches, and the envelope formed by sidebranches. 
\end{abstract}
\maketitle
\section{Introduction}
Branched patterns such as  diffusion-limited aggregation (DLA) and dendrite appear in various diffusion fields under strongly nonequilibrium conditions \cite{rf:1,rf:2,rf:3}. Dendrites with rough surfaces have been intensively studied by theoretical analyses, numerical simulations, and experiments \cite{rf:4,rf:5}. 
Two-dimensional DLA and dendrites were studied fairly well theoretically and experimentally; however, three-dimensional dendrites have not been sufficiently studied.  Huang and Glicksman studied dendritic tips and sidebranch structures of succinonitrile\cite{rf:6,rf:7}. They found the law $v\rho^2=$const for a dendritic tip, where $v$ is the tip velocity and $\rho$ is the radius of curvature, and found that the spacing of neighboring dominant sidebranches increases as $\Delta z\sim z^{1.3}$, where $z$ is the distance from the dendritic tip and $\Delta z$ is the spacing between neighboring dominant sidebranches. The dendritic tip is well approximated by a paraboloid and the cross section is almost circular (axisymmetric) for succinonitrile. More detailed studies were performed in microgravity using images taken from the space shuttle \cite{rf:8}. Sidebranches can grow owing to thermal noises \cite{rf:9} and develop far from the dendritic tip. 
Li and Beckermann found various scaling laws for sidebranches in an experiment on succinonitrile \cite{rf:10}. The envelope formed by sidebranches obeys $X\propto z^{0.859}$ and the volume $V$ between the dendritic tip and $z$ obeys $V\propto z^{2.1}$.  The cross section of the dendritic tip is not always axisymmetric and a fin structure develops in some dendrites. Brener and Temkin studied such non-axisymmetric three-dimensional dendrites theoretically and found that the shape of the dendritic tip can be described by $z=a|x|^{5/3}$, where $x$ is the length scale of the fin structure perpendicular to the growth direction. It is different from the parabolic shape $z=a|x|^2$ \cite{rf:11,rf:12}.  Bisang and Bilgram confirmed the scaling law of $z=a|x|^{5/3}$ in an experiment on xenon dendrites \cite{rf:13}. Wittwer and Bilgram measured the envelope of sidebranches of xenon dendrites and estimated $X \sim 0.5\cdot z$ \cite{rf:14}. 

 On the other hand, we proposed a coupled map lattice model as a simple simulation method of generating various growth patterns such as DLA, DBM, and dendrites \cite{rf:15,rf:16}. 
The envelope shape of two-dimensional dendrites was studied using a coupled map lattice \cite{rf:17}. The height distribution of sidebranches was studied using a coupled map lattice and compared with a needle model \cite{rf:18}. The height distribution obeys the power law $p(h)\sim h^{-1.9}$. The power-law distribution of sidebranches was confirmed in an experiment on quasi-two-dimensional NH$_4$Cl crystals \cite{rf:19}. In this study, we construct three-dimensional dendrites using a coupled map lattice, and study the fractal dimension, $f(\alpha)$ spectrum,  envelope shape, and height distribution of the sidebranhches of three-dimensional dendrites.       
\section{Three-Dimensional Coupled Map Lattice}
Dendritic crystals grow from a supersaturated solution.  Assuming the cubic symmetry for such crystals, a coupled map lattice model on a cubic lattice is proposed. The coupled map lattice for the solution growth is composed of two processes. One is the diffusion of the solution:
\begin{eqnarray}
u_{n}^{\prime}(i,j,k)& =&u_{n}(i,j,k)+D\{u_{n}(i+1,j,k)+u_{n}(i-1,j,k)+u_{n}(i,j+1,k)\nonumber\\
& &+u_{n}(i,j-1,k)+u_n(i,j,k+1)+u_n(i,j,k-1)-6u_{n}(i,j,k)\},
\end{eqnarray}
where $u$, and $u^{\prime}$ denote the dimensionless concentration of the solution, $n$ is the number of steps, $(i,j,k)$ denotes the lattice point, and \(D\) is the diffusion constant.
The second step is the growth at the interface.  
The order parameter $m(i,j,k)$ is introduced at each lattice point to express the growth. The order parameter \(m(i,j,k)\) is set to be 0 at sites of the solution, and 1 at sites of the crystals. 
The order parameter $m(i,j,k)$ changes only at the interface between solution sites and crystal sites.  
The growth rules of $m$ and $u$ at interface sites are written as
\begin{eqnarray}
m_{n+1}(i,j,k) & = & m_{n}(i,j,k)+c (u_{n}^{\prime}(i,j,k)-u_c),\nonumber\\
u_{n+1}(i,j,k) & = & u_{n}^{\prime}(i,j,k)-c (u_{n}^{\prime}(i,j,k)-u_c),
\end{eqnarray}
where $u_c$ denotes the equilibrium concentration of the solution, and $c$ denotes the kinetic coefficient of deposition from the solution into the crystals at interface sites, which is proportional to the local supersaturation $u-u_c$. The conservation law of mass is satisfied in our coupled map lattice model during growth at the interface, that is, the concentration of the solution around crystal sites decreases as crystalization proceeds.  If $m_{n+1}(i,j,k)$ goes over a threshold $m_c\sim 1$, the $(i,j,k)$ site is set to be a site of crystals. The same processes are repeated.  The threshold $m_c$ is 1 for the simplest model. Randomness can be incorporated by setting the threshold $m_c$ as a random number between $1-\Delta<m_c<1+\Delta$. 
We assume that $u_c$ is 0 when the effect of surface tension is not considered. The effect of surface tension is incorporated in our model by counting the number of neighboring sites of crystals, $N$, for each interface site, and setting $u_c=\gamma(N-9)$. Here, the neighboring sites are sites denoted by $(i\pm 1,j\pm 1,k\pm 1)$ around the site $(i,j,k)$. 
For a flat surface, $N=9$ and $u_c=0$. For a tip site, $N=1$ and $u_c=-8\gamma<0$; then growth is suppressed at such a protruded site. 
\section{Fractal Dimension and $f(\alpha)$ Spectrum}  
We have performed numerical simulations of a coupled map lattice. 
A typical three-dimensional dendrite is shown in Fig.~1.
A cubic lattice of $[-L/2,L/2] \times [-L/2,L/2] \times [-L/2,L/2]$ is used in numerical simulation. Fixed boundary conditions are imposed as $u(i,j,k)=u_0$ at $i=\pm L/2,j=\pm L/2$ or $k=\pm L/2$.
A point seed is set at the center $(0,0,0)$ where $L=300$.
 The initial concentration is $u(i,j,k)=u_0=0.01$. The other parameters are $c=1,\gamma=0, D=0.15$, and $\Delta=0$.  
Figure~1(a) shows a 3D plot of the dendrite for $k>60$, and Fig.~1(b) shows the cross section at $k=0$. 
\begin{figure}[tbp]
\begin{center}
\includegraphics[height=4.cm]{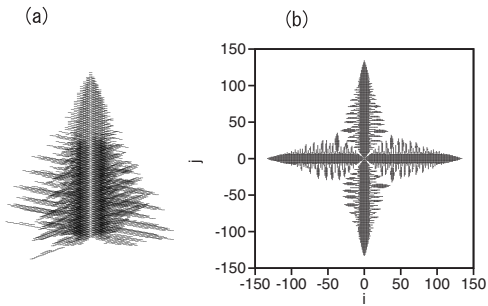}
\end{center}
\caption{(a) 3D plot of a three-dimensional dendrite for $u_0=0.01$. A region for $k>60$ is plotted. (b) Cross section of the three-dimensional dendrite at $k=0$.}
\label{fig1}
\end{figure}
\begin{figure}[tbp]
\begin{center}
\includegraphics[height=4.cm]{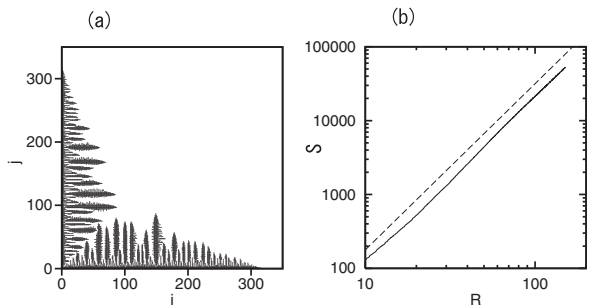}
\end{center}
\caption{(a) Cross section of a three-dimensional dendrite for $u=0.01$ and $\Delta=0.04$. (b) Relation of the radius $R$ of gyration and the number  of crystal sites, $S$. The dashed line denotes $S\sim R^{2.25}$. }
\label{fig2}
\end{figure}

Figure 2(a) shows the cross section of a larger three-dimensional dendrite for $i>0$ and $j>0$ at $k=0$ . The parameter values are  $u_0=0.01,\gamma=0,L=700$, and $\Delta=0.04$. During the dendritic growth, the total number  of crystal sites, $S_n$, is counted at a time step $n$, and  the radius  of gyration, $R_n$, at the same time step $n$ is calculated using $R_n^2=\sum_{i,j,k}r_{i,j,k}^2/S_n$, where $r_{i,j,k}$ is the distance between a crystal site $(i,j,k)$ and the origin. Figure 2(b) shows the relation of $R_n$ and $S_n$. The power law relation $S\sim R^{D}$ with $D\sim 2.25$ is observed. The exponent $D$  was evaluated using the data for $R>30$. $D$ is the fractal dimension of a three-dimensional dendrite. The fractal dimension of the three-dimensional DLA is $D\sim 2.5$, but there are few reports on the fractal dimension of three-dimensional dendrites. 
On the other hand, the fractal dimension of two-dimensional dendrites is $D\sim 1.5$,  which has also been checked using our two-dimensional coupled map lattice.

Next we calculated the $f(\alpha)$ spectrum for the growth probability \cite{rf:20}. 
We considered the fact that the concentration $u(i,j,k)$ at the interface site is proportional to growth probability. 
The partition function $Z_q(L)$  is calculated as
\begin{equation}
Z_q(L)=\sum \{P(i,j,k)\}^q,
\end{equation}
where $P(i,j,k)=u(i,j,k)/\sum u(i,j,k)$, and the summation is taken for all the interface sites.  The size  of the dendrite, $L_n$, grows with time; here $L_n$ is measured as the distance between the tip position and the origin. 
The generalized dimension $D_q$ is calculated using the partition function as
\begin{equation}
D_q=\frac{-1}{q-1}\lim_{L_n\rightarrow\infty}\frac{\ln Z_q(L_n)}{\ln L_n}.
\end{equation}
Numerically, $\lim_{L_n\rightarrow\infty}\{\ln Z_q(L_n)\}/\ln L_n$ was evaluated at the linear slope of the plot of $\ln Z_q(L_n)$ vs $\ln L_n$ for $L_n>30$. The $f(\alpha)$ spectrum is calculated using 
\begin{eqnarray}
\alpha(q)&=&\frac{d}{dq}\{(q-1)D_q\},\nonumber\\
f(\alpha(q))&=&q\alpha(q)-(q-1)D_q.
\end{eqnarray}
\begin{figure}[tbp]
\begin{center}
\includegraphics[height=4.cm]{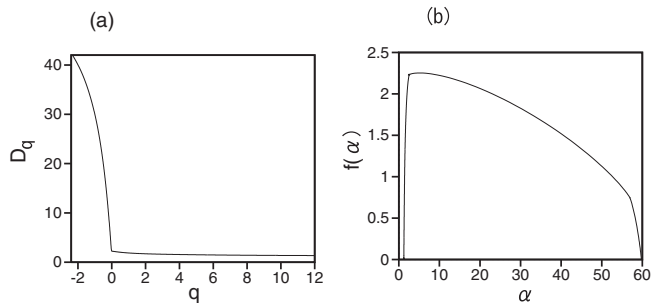}
\end{center}
\caption{(a) Relation of $q$ and $D_q$. (b) Relation of $\alpha$ and $f(\alpha)$  for $u=0.01$, $\gamma=0$, and $\Delta=0.04$.}
\label{fig3}
\end{figure}
Figure 3 shows the $D_q$ spectrum and $f(\alpha)$ spectrum of the dendrite at $u_0=0.01,\gamma=0,L=700$, and $\Delta=0.04$. 
The peak value of $f(\alpha)$ is 2.24, which is the fractal dimension $D_0$.
The information dimension $D_1$ is evaluated as $D_1=1.87$, and $D_{\infty}$ is evaluated as $D_{\infty}=1.34$. The minimum $\alpha$ is evaluated as $\alpha_{min}=1.24$. 
The $f(\alpha)$ spectrum is rather wide in comparison with that of the two-dimensional dendrite. The spectrum for $D(q)$ for $q<-2.4$ could not be accurately calculated, because a clear linear relation of $\ln Z_q(L_n)$ and $\ln L_n$ was not obtained or the data scattered widely for small values of $q$. Then, the $f(\alpha)$ spectrum is not accurately calculated for $\alpha>56$. We have also calculated the $f(\alpha)$ spectrum of dendrites at other parameter values such as $u_0=0.005, 0.02$ and  $\Delta=0,0.02$, but the $f(\alpha)$ spectrum does not change markedly.
\section{Envelope of Sidebranches}
We have studied three-dimensional dendrites with surface tension using the coupled map lattice model by incorporating the parameter $\gamma$.
\begin{figure}[tbp]
\begin{center}
\includegraphics[height=3.5cm]{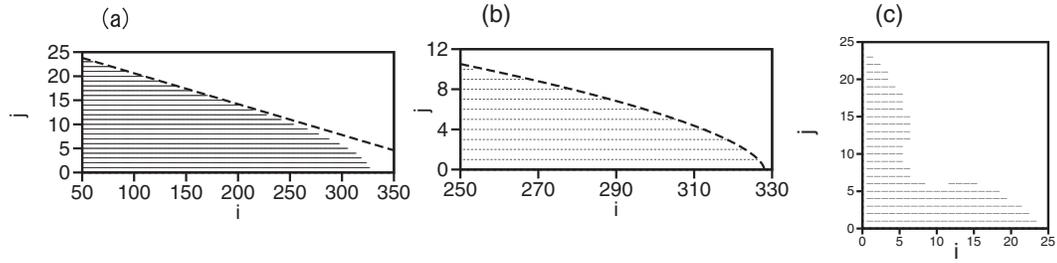}
\end{center}
\caption{(a) Cross section of a three-dimensional dendrite in the $(i,j)$ space at $k=0$ for $i>50$. The parameter $\gamma$ is $4\times 10^{-5}$. The dashed line is $j=27-0.064i$.  
(b) Cross section of a three-dimensional dendrite in the $(i,j)$ space for $i>250$. The dashed curve is $j=0.77\cdot (328-i)^{0.6}$. (c) Cross section of a three-dimensional dendrite in the $(j,k)$ space at $i=50$.}
\label{fig4}
\end{figure}
\begin{figure}[tbp]
\begin{center}
\includegraphics[height=3.5cm]{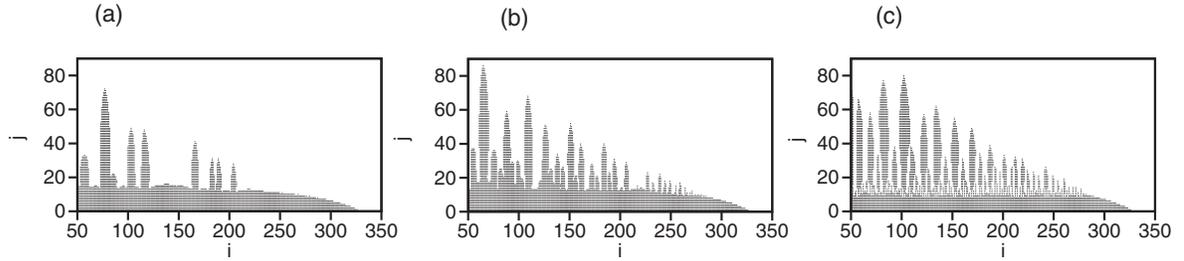}
\end{center}
\caption{Cross section of a three-dimensional dendrite in the $(i,j)$ space at $k=0$ for $i>50$. The parameter $\gamma$ is (a) $2.5\times 10^{-5}$, (b) $1.5\times 10^{-5}$, and (c) $0$.}
\label{fig5}
\end{figure}
Figure 4(a) shows a cross section of a dendrite at $k=0$ and $i>50$ for $c=0.8$, $\gamma= 4\times 10^{-5}, u_0=0.005$, and $\Delta=0$. A smooth interface appears owing to the surface tension effect. The interface is approximated as $j=27-0.064i$ for $i<200$. Figure 4(b) shows the magnification of the tip region. Near the tip region, the interface is approximated by $j=0.77\cdot (328-i)^{0.6}$. This is consistent with the scaling law $x\sim z^{0.6}$ by Brener, because  $z$ is the distance from the dendritic tip and $x$ is the height of the fin structure in the notation by Brener, and $z$ and $x$  respectively correspond to $328-i$ and $j$ in our coupled map lattice. 
Figure 4(c) shows a cross section in the $(j,k)$ space at $i=50$. A two-dimensional dendrite appears in this cross section.  The ratio of the tip velocity in the $j$-direction of this two-dimensional dendrite and the tip velocity in the 
$i$-direction of the three-dimensional dendrite shown in Fig.~4(a) is equal to the slope 0.064 of the dashed line in Fig.~4(a).  

\begin{figure}[tbp]
\begin{center}
\includegraphics[height=4.cm]{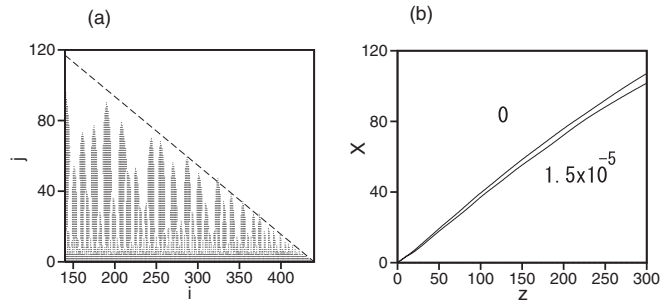}
\end{center}
\caption{(a) Snapshot pattern at $u_0=0.005$ and $\gamma=0$. (b) Envelopes of sidebranches for $\gamma=0$ and $\gamma=1.5\times 10^{-5}$.}
\label{fig6}
\end{figure}
The smooth interface becomes unstable for $\gamma<2.7\times 10^{-5}$.  Figure 5 shows snapshot patterns for $\gamma=2.5\times 10^{-5},1.5\times 10^{-5}$, and 0 for $u_0=0.005$ and $\Delta=0$. Sidebranches appear in a sporadic manner far from the dendritic tip  at $\gamma=2.5\times 10^{-5}$. The number density of sidebranches increases as $\gamma$ decreases.  We have constructed an envelope of sidebranches along the $i$ axis by connecting the tip positions $j_t(i)$ of sidebranches that are taller than any tip positions $j_t(i^{\prime})$ at $i^{\prime}$, satisfying  $i^{\prime}>i$. We have performed many numerical simulations and constructed an average shape of the envelope of sidebranches. 
In these numerical simulations, the random threshold of  $\Delta=0.04$ is assumed only on the three lines of $(i,0,0),(0,j,0)$, and $(0,0,k)$ passing through the origin. Figure 6(a) shows a snapshot pattern for $u_0=0.005$ and $\gamma=0$. 
The dashed line is $j=0.39\cdot(440-i)$.
Figure 6(b) shows average shapes of the envelopes of sidebranches at $\gamma=0$ and $\gamma=1.5\times 10^{-5}$, where $z$ is the distance from the dendritic tip at $j=k=0$, and $X$ is the height of the envelope at $z$.  The envelopes are approximated at $X\sim 0.39 \cdot z$ for $\gamma=0$ and $X\sim 0.37 \cdot z$ for $\gamma=1.5\times 10^{-5}$ for $z<150$, and the envelopes bend slightly downward as $X\sim z^{0.9}$ for $z>150$. These results are consistent with the experimental results.  
\section{Competitive Growth among Sidebranches}
Sidebranches compete with each other via the diffusion field. 
When a sidebranch grows slightly faster than the neighboring sidebranches, the taller sidebranch grows even faster because the sidebranch is in contact with a denser diffusion field. In a previous paper, we performed a numerical simulation of the competitive growth of two sidebranches using a two-dimensional coupled map lattice\cite{rf:17}. We have found that the growth velocity of a shorter sidebranch decays in an exponential manner, and that decay rate decreases with the power law of the distance  between the two sidebranches, $l$. We have performed a similar numerical simulation in three dimensions. We have used a rectangular box of $L_x\times L\times L$ for numerical simulation, and periodic boundary conditions are imposed in the $x$-direction. As an initial condition, we have set two seeds of crystals of different sizes at $i=i_1=L_x/2-(l+1)/2$ and $i=i_2=L_x/2+(l+1)/2$. Figure 7(a) shows a snapshot at $n=150000$ for $u_0=0.005, \gamma=0,\Delta=0, L=280,L_x=22$, and $l=11$. 
The right branch at $i=17$ grows faster and wins the competition, because the initial seed is slightly larger.  
Figure 7(b) shows the time evolution of the tip positions $j_{t1}$ and $j_{t2}$. The tip position $j_{t1}$ of the left branch stops to grow because the right branch dominates the diffusion field.  Figure 7(c) shows a semilogarithmic plot of the growth velocity $v_1$ of the left branch as a function of the difference $\Delta=j_{t2}-j_{t1}$ of the tip positions for $l=11$ and 19. 
The velocity $v_1$ decays slower than an exponential decay. The dashed curves are fitting curves: $0.0012\exp(-0.75\Delta^{0.45})$ for $l=11$ and $0.0004\exp(-0.15\Delta^{0.65})$ for $l=19$. It is different from the exponential decay observed in two-dimensional dendrites in our previous paper.  We do not understand the reason for the stretched exponential decay, however, the slower decay implies that the suppression effect by the longer branch is weaker in three dimensions. This is probably because the shorter branch can grow slightly owing to the diffusion of $u$ in the $k$-direction. 

\begin{figure}[tbp]
\begin{center}
\includegraphics[height=4.5cm]{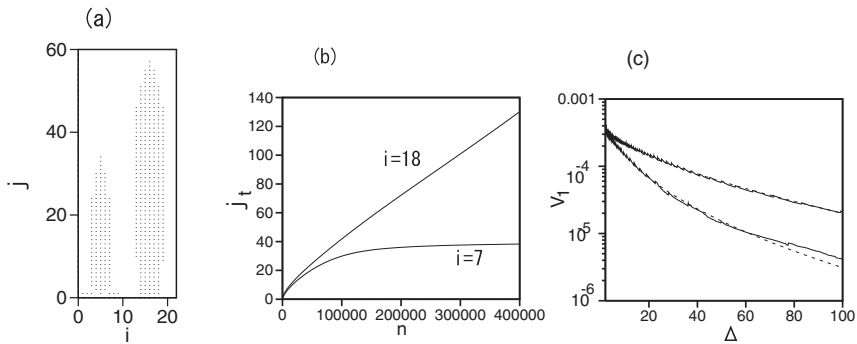}
\end{center}
\caption{(a) Competition of two branches at $u_0=0.005$ The system size is $22\times 280\times 280$. (b) Time evolution of the two tip positions at $i=7$ and $18$. (c) Relation of $v_1$ and the difference between the two tip positions for $l=11$ and 19. 
}
\label{fig7}
\end{figure}
\begin{figure}[tbp]
\begin{center}
\includegraphics[height=4.cm]{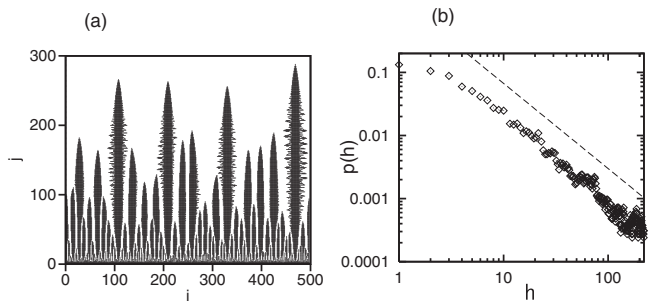}
\end{center}
\caption{(a) Snapshot pattern at $u_0=0.005$ and $\gamma=0$ grown from a linear seed at $j=0$.
(b) Size distribution $p(h)$ of branch size $h$.  The dashed line denotes a power law of exponent 1.35.}
\label{fig8}
\end{figure}
Next, we have performed a numerical simulation from a linear seed at $j=0$ and $k=0$ to study the competitive growth of many sidebranches. 
Figure 8(a) shows a snapshot pattern at $k=0$ for $u_0=0.005,\gamma=0$, and $\Delta=0$. The initial $m(i,j,k)$ at interface sites around the linear seed is a random number between 0 and 0.2. 
Many branches grow and comptete with each other. Several large branches survive and grow.  Finer-scale sidebranches develop on the surviving large branches.  The growth of other small branches is suppressed owing to the larger sidebranches. 
Figure 8(b) shows the size distribution $p(h)$ of branches on a double-logarithmic scale. The size distribution was constructed from 22 branch patterns in the $i-j$ space and $i-k$ space obtained from 11 random initial conditions. 
The approximate curve is the power law of $p(h)\sim 1/h^{1.35}$.  The exponent of this power law was 1.9 in two-dimensional coupled map lattices.  The exponent in our three-dimensional coupled map lattice is rather small, which might be related to the fact in Fig.~7 that the competition is weak in three dimensions.  

\section{Conclusions}
We have performed numerical simulations of three-dimensional dendrites using coupled map lattices. We have succeeded in generating well-developed three-dimensional dendrites. We have calculated the fractal dimension and $f(\alpha)$ spectrum.  We have checked the scaling law $x\sim z^{3/5}$ of the smooth tip region by Brener by incorporating a surface tension effect, and found a transition from a smooth surface to a well-developed dendrite. 
Furthermore, we have studied competitive dynamics among sidebranches, and found some difference of what from two-dimensional dendrites.   

\end{document}